\documentclass[a4paper,letter]{jmc}
\usepackage{graphicx}
\usepackage{xspace,colortbl}
\usepackage{amsfonts}
\usepackage{amsmath}
\usepackage{amssymb}
\usepackage{xtab}
\usepackage{rotating}
\usepackage{longtable,geometry}
\usepackage[english]{babel}
\usepackage[utf8]{inputenc}
\usepackage[babel]{csquotes}
\usepackage{listings}
\usepackage[usenames,dvipsnames,svgnames,table]{xcolor}
\usepackage{multicol}
\usepackage{float}
\usepackage{centernot}
\usepackage[colorlinks=true,linkcolor=Black,urlcolor=MidnightBlue,citecolor=MidnightBlue]{hyperref} 

\newcommand{\JCTCformat}[4]{{\it #1} {\bf #2}, {\it #3}, {#4}.}

\newcommand{\Refpcc}[4]{\JCTCformat{#1}{#2}{#3}{#4}}

\newcommand{\jcp}[3]{\Refpcc{J. Chem. Phys.}{#1}{#2}{#3}}

\newcommand{\jmathphys}[3]{\Refpcc{J. Math. Phys.}{#1}{#2}{#3}}

\newcommand{\molphys}[3]{\Refpcc{Mol. Phys.}{#1}{#2}{#3}}

\newcommand{\revmodphys}[3]{\Refpcc{Rev. Mod. Phys.}{#1}{#2}{#3}}

\newcommand{\overbar}[1]{\mkern 3mu\overline{\mkern-3mu#1\mkern-1mu}\mkern 1mu}

\usepackage{stmaryrd} 

\DeclareFontFamily{U}{mathb}{\hyphenchar\font45}
\DeclareFontShape{U}{mathb}{m}{n}{
      <5> <6> <7> <8> <9> <10> gen * mathb
      <10.95> mathb10 <12> <14.4> <17.28> <20.74> <24.88> mathb12
      }{}
\DeclareSymbolFont{mathb}{U}{mathb}{m}{n}
\DeclareMathSymbol{\dlsh}{3}{mathb}{"EA}

\begin{document}

\begin{frontmatter}

\title{A Conjecture on Antisymmetrized Geminal Power Wavefunctions}

\author{Patrick Cassam-Chena\"{\i}} 

\address{Universit\'e C\^ote d'Azur, LJAD, UMR 7351, 06100 Nice, France}

\begin{abstract}
We conjecture that ``Antisymmetrized Geminal Power'' wave functions, and, in particular, those of extreme type  in Coleman's terminology (i.e. with all geminal coefficients equal), can always be rewritten as antisymmetrized products of geminals orthogonal to each other. We prove this conjecture in simple cases and provide numerical evidence that it holds true in more complicated examples. Establishing the validity of this conjecture is important as it questions the physical interpretation of AGP wavefunctions. 
\end{abstract}
\end{frontmatter}

Keywords:\\
Antisymmetrized Geminal Power, Superconductivity, 

Suggested running head:
Conjecture on AGP

All correspondance to be send to P. Cassam-Chena\"{\i},\\
cassam@unice.fr,\\
tel.: +33 4 92 07 62 60,\\
fax:  +33 4 93 51 79 74.

%

\newpage
\section{Introduction}

 Antisymmetrized Geminal Power (AGP) wave functions have appeared in the 1950's in superconductivity theory and were called Schafroth-condensed pair wave functions in this context~(\cite{Blatt1964}, p.169). Later, it was noticed~(\cite{Blatt1964}, p.183) that they correspond to number projected Bardeen-Cooper-Schrieffer (BCS) wave functions~\cite{Bardeen57}, and they have attracted a lot of attention from the mathematical chemistry and quantum chemistry communities in the 1960's \cite{Coleman63,Coleman65,Bratos65}. In the past twenty years, they have become a common factor of test functions in quantum Monte-Carlo calculations~\cite{Casula2003}. More recently, several authors have advocated their use as reference states for more involved post-treatments of electron correlation~\cite{Neuscamman2013-CJAGP,Neuscamman2016-JCTC,Neuscamman2016-CJAGP,Neuscamman2012,Neuscamman2013-JAGP,Zhao2016-JCTC,Kawasaki2016,Henderson2019,Henderson2020,Khamoshi21} and the AGP ansatz has even been ported to quantum computers~\cite{Khamoshi20}.
 
 Of particular importance are the AGP of extreme type, in Coleman's terminology, which have many remarkable properties, such as having reduced density matrices (RDM) whose largest eigenvalue achieves the maximal possible value for a $2n$-Fermion wavefunction within a $2m$-dimensional one-particle Hilbert space~\cite{Coleman63}. It is often argued that a large 2-particle RDM (2-RDM) eigenvalue is a ``key to the understanding of superconductivity''~\cite{Coleman65} as it accompanies the onset of off-diagonal-long-range-order in the 2-RDM. It is usually believed that the seemingly ``bosonic character'' of the wave function structure as a power of a Fermion pair, is essential to the coherent pairing in the natural geminal associated to the largest 2-RDM eigenvalue.
 
 The main purpose of this article is to propose a conjecture, which, if true, could lead to a revision of such belief. In the next section, we focus on AGP of extreme type. The conjecture is proved in two extreme cases, namely $n=m$, and $n=2$, $m$ arbitrary. It has been verified numerically for many values of $m$ up to $n=5$. In the last section, we consider the possible extension of the conjecture and conclude with a hint for a demonstration.  
 
 \section{Conjecture for AGP of extreme type}
 
Let $\mathcal{H}$ be a $2m$-dimensional Hilbert space. We note $(\varphi_1,\ldots,\varphi_m,\overbar{\varphi}_1,\ldots,\overbar{\varphi}_m)$ an orthonormal basis set of this space. Consider the two-fermion wave function i.e. the "geminal", $g$ defined by:
\begin{equation*}
 g=\left(\frac{1}{n!\sqrt{\binom{m}{n}}}\right)^{\frac{1}{n}}\sum\limits_{i=1}^m \varphi_i\wedge\overbar{\varphi}_i
\end{equation*}
and its $n^{\mathit{th}}$ antisymmetrized power:
\begin{equation*}
 g^{\wedge n}=g\wedge\cdots\wedge g = \frac{1}{\sqrt{\binom{m}{n}}}\:\sum\limits_{1\leq i_1<i_2<\cdots<i_n\leq m} \varphi_{i_1}\wedge\overbar{\varphi}_{i_1}\wedge\cdots\wedge\varphi_{i_n}\wedge\overbar{\varphi}_{i_n}
\end{equation*}
The coefficients of $g$  are all equal, which makes $g^{\wedge n}$ an Antisymetrized Geminal Power (AGP) of extreme type in Coleman's denomination \cite{Coleman63,Coleman65}. The normalization of $g$ has been chosen so that $g^{\wedge n}$ be normalized to $1$.


We conjecture that there always exists $n$ geminals $g_1,\ldots,g_n$ of the form:
\begin{equation*}
 g_k=\sum_{i=1}^m c^k_i\,\varphi_i\wedge\overbar{\varphi}_i+\sum_{i\neq j} d^k_{i,j}\,(\varphi_i\wedge\overbar{\varphi}_j+\varphi_j\wedge\overbar{\varphi}_i)
\end{equation*}
such that $g_1\wedge\cdots\wedge g_n=g^{\wedge n}$ and $g_k\perp g_l$ i.e. $\langle g_k | g_l\rangle=0$ for all $k\neq l$.

The conjecture is always true in the case $m=n$, since, as is well-known, $g^{\wedge n}$ is a single Slater determinantal function in this case, that can be rewritten as:
\begin{equation*}
 g^{\wedge n}= \varphi_{1}\wedge\overbar{\varphi}_{1}\wedge\cdots\wedge\varphi_{n}\wedge\overbar{\varphi}_{n},
\end{equation*}
so, it suffices to set: $\forall i,\ g_i=\varphi_i\wedge\overbar{\varphi}_i$. Hereafter, we will assume $m>n$. 

The conjecture is also true for $n=2$ and arbitrary $m$. A solution is for example: 
\begin{equation}
 g_1=\sum\limits_{i=1}^{m-1} \varphi_i\wedge\overbar{\varphi}_i+(1+\sqrt{m})\, \varphi_m\wedge\overbar{\varphi}_m \, ,
 \label{n=2-g1}
\end{equation}
\begin{equation}
 g_2=\frac{1}{\sqrt{2m(m-1)}}\left(\sum\limits_{i=1}^{m-1} \varphi_i\wedge\overbar{\varphi}_i+(1-\sqrt{m})\, \varphi_m\wedge\overbar{\varphi}_m\right).
  \label{n=2-g2}
\end{equation}

These two cases suggest to limit the search of $g_k$'s to seniority $0$ geminals i.e. to set $\forall i, j, k, d^k_{i,j}=0$ in the first place.

The $g_k$'s have to satisfy two types of constraints:\\
i) Orthogonality constraints,
\begin{equation}
 \forall k\neq l, \ \sum_{i=1}^m c^k_i c^l_i =0 \ .
 \label{orthogonality}
\end{equation}
ii) ``Coefficient'' constraints, obtained by identification of the $2n$-Fermion expansion coefficients in the induced basis on the $2n$-particle space.
The coefficients of $g_1\wedge\cdots\wedge g_n$ involved permanents, noted "$|\quad|_+$". The constraints are expressed as follows:
\begin{equation}
 \forall (i_1, i_2, \cdots , i_n)\ \mathit{such\ that\ } 1\leq i_1<i_2<\cdots<i_n\leq m, \ \left|\begin{array}{cccc}
 c_{i_1}^1 & c_{i_1}^2 & \cdots & c_{i_1}^n \\
 c_{i_2}^1 & c_{i_2}^2 & \cdots & c_{i_2}^n \\
 \vdots & \vdots & \ddots & \vdots \\
 c_{i_n}^1 & c_{i_n}^2 & \cdots & c_{i_n}^n \\
 \end{array}\right|_+=\frac{1}{\sqrt{\binom{m}{n}}}\ .
 \label{coef-cond}
\end{equation}

We thus have $m\times n$ parameters (the $c^k_i$'s), for $\binom{n}{2}+\binom{m}{n}$ orthogonality contraints and coefficient constraints, which  \textit{a priori} are not necessarily independent.\\

By borrowing from the case $n=2$, the special structure of its special solution given in Eqs. (\ref{n=2-g1}) and (\ref{n=2-g2}), we further constrain  for each geminal $k$, the coefficients $c^k_i$'s  to be all equal for $i\in\{1, \cdots,m-n+1\}$. When $j$ indices among $i_1,i_2, \ldots ,i_n$ belong to $\{m-n+2, \cdots ,m\}$, this reduces the number of distinct coefficient conditions to $\binom{n-1}{j}$,  and since $j$ takes all values in $\{0,\cdots ,n-1\}$, this leads to $2^{n-1}$ independent coefficient conditions altogether, from Eq.(\ref{coef-cond}). It is also convenient to enforce that all the $g_k$'s but $g_1$ be normalized to $1$, adjusting the normalization of the latter to normalize the whole antisymmetrized product. All this simplifies our search for $n>2$ solutions: the number of free parameters is reduced to $n^2$, and the total number of constraints to $\frac{(n-1)(n+2)}{2} + 2^{n-1}$, both independent of $m$.

Even within these restrictions, we have obtained numerical solutions for $n\leq 5$ and all values of $m$ we have tried.  For example, for $n=5$ and $m=30$, we let the reader verify that the following set of coefficients satisfies all the required conditions to an accuracy better than $10^{-19}$.

\begin{tiny}
\begin{center}
\begin{xtabular}{c|ccccc}
k           & $c^k_1$ to $c^k_{26}$       &    $c^k_{27}$               & $c^k_{28}$  & $c^k_{29}$ &  $c^k_{30}$ \\
1& 0.57379225208684241102& -0.97314596644759151711& -0.52192547694758039956& -0.32715721668013876005& -4.0193331890144964655\\
2& 0.083540800636536504358& 0.68201678831451321700& -0.52573644857790835248& -0.16059374532554361624& 0.22629236931528920666\\
3& 0.073898881274917732634& 0.23582864474564844539& 0.79315210502272865074& -0.38988528747375119946& 0.14593454091188107943\\
4& 0.065251642336844446912& 0.14043162229985210050& 0.22271804741572421423& 0.89928944041885804513& 0.10607514150664062435\\
5& 0.095503558806451137654& -0.65050104401723396498& -0.18559956378216714883& -0.096786316693051254828&0.54395719426736260336\\
 \hline
\end{xtabular}
\end{center} 
\end{tiny}

In fact, it seems easy to find a solution of this type, even though the number of constraints, $30$, is larger than the number of parameters, $25$, in the case $n=5$, whatever the value of $m$ might be. 

\section{Generalizations and conclusion}

So far, we have focused on AGP of extreme type, since their very special properties tend to suggest that they are the most difficult AGP to approach with orthogonal geminals. However,  the conjecture might also hold true for general AGP, that is those constructed as an antisymmetrized power of a geminal of the form:
\begin{equation*}
 g=\sum\limits_{i=1}^m a_i\, \varphi_i\wedge\overbar{\varphi}_i.
\end{equation*}

For the case $n=2$ our particular solution, Eqs. (\ref{n=2-g1}) and (\ref{n=2-g2}), can easily be adapted:
\begin{equation*}
 g_1=\sum\limits_{i=1}^{m-1} a_i\varphi_i\wedge\overbar{\varphi}_i+\left(a_m+\sqrt{\sum\limits_{i=1}^m |a_i|^2}\right)\, \varphi_m\wedge\overbar{\varphi}_m,
\end{equation*}
\begin{equation*}
 g_2=\sum\limits_{i=1}^{m-1} a_i\varphi_i\wedge\overbar{\varphi}_i+\left(a_m-\sqrt{\sum\limits_{i=1}^m |a_i|^2}\right)\, \varphi_m\wedge\overbar{\varphi}_m,
\end{equation*}
gives
\begin{equation*}
 g_1 \wedge g_2 = \sum\limits_{1\leq i<j \leq m} 2\, a_i a_j \varphi_i\wedge\overbar{\varphi}_i\wedge \varphi_j\wedge\overbar{\varphi}_j = g^{\wedge 2}
\end{equation*}
while  $\langle g_1 | g_2 \rangle=0$.

It is even tempting to go beyond AGP wave functions and state that the set of antisymmetrized product of orthogonal geminal  is dense in the set of antisymmetrized product of any geminals, i.e. given $n$ geminals $g_1', g_2', \ldots , g_n'$,
\begin{equation*}
\forall \varepsilon>0,\, \exists\ g_1, g_2, \ldots , g_n, \mathit{such\, that}\ \forall i\neq j, \langle g_i | g_j \rangle=0\ \mathit{and}\  || g_1\wedge\cdots\wedge g_n - g_1'\wedge\cdots\wedge g_n' || < \varepsilon.
\end{equation*}

In Eq.(\ref{orthogonality}), we have implicitly assumed real coefficients. However, we can of course consider the complex case, which can provide us more freedom. The scalar product has to be replaced by the Hermitian product and the orthogonality constraints become: 
\begin{equation}
 \forall k\neq l, \ \sum_{i=1}^m \bar{c}^k_i c^l_i =0,
 \label{hermitian-prd}
\end{equation}
where $\bar{c}^k_i$ denotes the complex conjugate of $c^k_i$.
On another hand, in the complex field, it is natural to invoke Hilbert's Nullstellensatz theorem to prove the existence of a solution to a set of polynomial equations. So, for each complex parameter, we set  $c^k_i= X^k_i + \imath Y^k_i$ and $\bar{c}^k_i= X^k_i - \imath Y^k_i$. Then, we substitute these expressions for all $i$ and $k$ into the constraints. We obtain a set of polynomial expressions in the indeterminates $(X^k_i)_{i,k}$ and $(Y^k_i)_{i,k}$. The orthogonality constraints, Eq.(\ref{hermitian-prd}), give homogeneous polynomials of degree 2,
\begin{equation}
 \forall k\neq l, \ \sum_{i=1}^m (X^k_i\cdot X^l_i + Y^k_i\cdot Y^l_i) + \imath (X^k_i\cdot Y^l_i - Y^k_i\cdot X^l_i).
 \label{ortho-poly}
\end{equation}
The coefficient constraints give polynomials that have only monomials of degree $n$ plus a constant i.e. of the form, for AGP of extreme type:
\begin{equation}
 \sum\limits_{\sigma\in\mathcal{S}_n} \prod\limits_{k=1}^n (X^k_{\sigma(i_k)}+ \imath Y^k_{\sigma(i_k)})-\frac{1}{\sqrt{\binom{m}{n}}},
 \label{coef-poly}
\end{equation}
where the sum extends over permutations of $n$ elements.
It can be easily seen that the constant $1$ is not in the ideal generated by these polynomials: if $1$ was in the ideal, it could be expressed as a linear combination of products of these polynomials, where there should be at least a non zero-coefficient in front of a power of a ``coefficient constraint'' polynomial of expression (\ref{coef-poly}), because the ``orthogonality constraint'' polynomials of expression (\ref{ortho-poly}), have no constant term. But then, the homogenous monomials of degree $n$ of this term would not possibly cancel out with some other terms, because, the polynomials of expression (\ref{ortho-poly}) have necessarily indeterminate subscript coming in pairs of identical indices, the others polynomials from expression (\ref{coef-poly}) give homogeneous monomials of degree $n$ that are all distinct, and the powers of the same polynomial form a free set. So, since $1$ is not in the ideal, the weak Hilbert's Nullstellensatz theorem applies and insures that all the polynomials have common zeros. However, it remains to prove that there are common zeros that are real, otherwise the zeros of the polynomials in (\ref{ortho-poly}) will not be suitable solutions of the orthogonality constraints Eq.(\ref{hermitian-prd}).

We leave a complete proof of the conjecture to the sagacity of the readers. We hope that the cases exhibited in this work up to $n=5$, that is $10$ molecular electrons for example in quantum chemistry, will already stimulate interest. 

\section*{ACKNOWLEDGEMENTS}
We acknowledge discussions with Prof. B. Mourrain, and contributions from two former students, T. Perez who performed an extended study of the case $n=2$ for AGP of extreme type and D. Bergeault who worked on using  Hilbert's Nullstellensatz theorem for proving the conjecture.

\section*{\begin{large}Declarations                      \end{large}}
\textbf{Ethical Approval}\\
Not applicable

\textbf{Competing interests}\\
None

\textbf{Authors' contributions}\\
Not applicable

\textbf{Funding}\\
None

\textbf{Availability of data and materials}\\
Not applicable

\end{document}